\documentstyle[12pt]{article}
\begin{document}

\def \inbar{\vrule height1.5ex width.4pt depth0pt}
\def \C{\relax\hbox{\kern.25em$\inbar\kern-.3em{\rm C}$}}
\def \R{\relax{\rm I\kern-.18em R}}
\newcommand{\Z}{\ Z \hspace{-.08in}Z}
\newcommand{\be}{\begin{equation}}
\newcommand{\ee}{\end{equation}}
\newcommand{\bea}{\begin{eqnarray}}
\newcommand{\eea}{\end{eqnarray}}
\newcommand{\nn}{\nonumber}
\renewcommand{\ll}{\left[ }
\newcommand{\rr}{\right] }
\newcommand{\kt}{\rangle}
\newcommand{\br}{\langle}
\newcommand{\lll}{\left( }
\newcommand{\rrr}{\right)}
\newcommand{\dagg}{\dagger}
\newcommand{\rdel}{\stackrel{\rightarrow}{\partial}}
\newcommand{\ldel}{\stackrel{\leftarrow}{\partial}}
\newcommand{\pbra}{[\hspace{-.3mm}[}
\newcommand{\pket}{]\hspace{-.3mm}]}

\title{Parabose -- Parafermi Supersymmetry}
\author{Ali Mostafazadeh \\ ~\\
Institute for Studies in Theoretical Physics and Mathematics,\\
P.~O.~Box 19395-5746, Tehran, Iran, and\\
Department of Physics, Sharif University of Technology,\\
P.~O.~Box 11365-9161, Tehran, Iran.\thanks{E-mail:
``alim@netware2.ipm.ac.ir'', Fax: (98-21)228-0415.}}
\maketitle

\baselineskip=24pt

\begin{abstract}
The ($p=2$) parabose -- parafermi supersymmetry is studied in general terms.
It is shown that the algebraic structure of the ($p=2$) parastatistical
dynamical variables allows for (symmetry) transformations which mix
the parabose and parafermi coordinate variables. The example of a simple
parabose -- parafermi oscillator is discussed and its symmetries investigated.
It turns out that this oscillator possesses two parabose -- parafermi
supersymmetries. The combined set of generators of the symmetries forms the
algebra of supersymmetric quantum mechanics supplemented with an additional
central charge. In this sense there is no relation between the parabose --
parafermi supersymmetry and the parasupersymmetric quantum mechanics.
A precise definition of a quantum system involving this type of parabose --
parafermi supersymmetry is offered, thus introducing ($p=2$) Supersymmetric
Paraquantum Mechanics. The spectrum degeneracy structure of general ($p=2$)
supersymmetric paraquantum mechanics is analyzed in detail.
The energy eigenvalues and eigenvectors for the parabose -- parafermi
oscillator are then obtained explicitly. The latter confirms the validity
of the results obtained for general supersymmetric paraquantum mechanics.
\end{abstract}

\newpage

\section{Introduction}
In a preceding article \cite{p10}, an attempt is made to simplify the
study of the algebraic structure of dynamical systems involving
($p=2$) parabose and parafermi variables. The approach presented in
\cite{p10} is aimed to facilitate the analysis of systems possessing
parabose -- parafermi supersymmetry, thus providing the necessary framework
for investigating the relation between the conventional parastatistics
of Green \cite{green} and the more recent developments of parasupersymmetric
quantum mechanics \cite{r-s,b-d,p8,p9}. More specifically the purpose
of the present article is to answer to the question:
        \begin{itemize}
        \item[]
        {\em Is parabose -- parafermi supersymmetry the same as
        parasupersymmetry?}
        \end{itemize}
One should note that the so-called ``parasupersymmetric oscillators''
studied in the literature, e.g., \cite{r-s,b-d}, are constructed using
some specific matrix representation of the parafermi operators. The
analysis presented in this paper does not restrict to matrix representations
and treats the parafermi and parabose operators (variables) as fundamental
mathematical objects.

The paper is organized as follows: In Sec.~2, the main results of \cite{p10}
are quoted and the possibility of the existence of parabose -- parafermi
(supersymmetry) transformations is investigated. In Sec.~3, the analogy
between the ($p=2$) parabose -- parafermi supersymmetry and the ordinary
bose -- fermi supersymmetry is discussed. The example of the supersymmetric
oscillator is then reviewed and the ($p=2$) parabose -- parafermi
oscillator is introduced by analogy. In Sec.~4, the parabose -- parafermi
supersymmetries of the oscillator are studied. In Sec.~5, the super Lie
algebra of the symmetries of the oscillator is used to define the
notion of {\em ($p=2$) Supersymmetric Paraquantum Mechanics}. This section
also offers a detailed treatment of the degeneracy structure of general
($p=2$) supersymmetric paraquantum mechanics. Sec.~6 is devoted to an
analysis of the energy eigenstates and the spectrum degeneracy
of the parabose -- parafermi oscillator. Here, the explicit form of
a complete set of energy eigenstate vectors is obtained. Sec.~7
includes the conclusions.

For brevity we shall use the notation $\pi b$, $\pi f$, $\pi SUSY$ for
{\em ($p=2$) parabose, parafermi,} and {\em parabose -- parafermi
supersymmetry}, and abbreviations SQM, PSQM, SPQM for {\em supersymmetric
quantum mechanics}, {\em parasupersymmetric quantum mechanics}, and
{\em supersymmetric paraquantum mechanics}, respectively. We shall follow
Einstein summation convention of summing over repeated indices throughout
the paper, unless otherwise indicated.

\section{Algebraic Structure of Classical $\pi SUSY$}
In this section, first we recall the constructions developed in \cite{p10}.

The algebra of the creation $a_k^{\mu\dagger}$ and annihilation operators
$a_k^{\mu}$ for the ($p=2$) $\pi b$ ($\mu=0$) and $\pi f$ ($\mu=1$)
variables is given by:
        \bea
        a_k^{\mu}&=&\sum_{\alpha=0}^1\zeta^{\alpha\mu}_k\;,
        \label{e1}\\
        \theta_{k 1}^{\alpha\mu}&:=& \sqrt{\frac{\hbar}{2}}
        (\zeta_k^{\alpha\mu}+
        \zeta_k^{\alpha\mu\dagger})\;,
        \label{e2} \\
        \theta_{k 0}^{\alpha\mu}&:=& -i\sqrt{\frac{\hbar}{2}}
        (\zeta_k^{\alpha\mu}-\zeta_k^{\alpha\mu\dagger})\;,
        \label{e3}\\
        \pbra \theta_{im}^{\alpha\mu},\theta_{jn}^{\beta\nu}\pket&:=&
        \hbar\delta_{ij}\delta^{\alpha\beta}
        [i(1-\mu)(1-\nu)\epsilon_{mn}+\mu\nu\delta_{mn}]\;,
        \label{e4}
        \eea
where $\zeta$'a are the Green components of $a$'s \cite{green}, $\alpha,
\beta,\mu,\nu=0,1$, $m,n=1,2$, and $\pbra~,~\pket$ is the parabraket:
        \be
        \pbra \theta_{im}^{\alpha\mu},\theta_{jn}^{\beta\nu}\pket
        := \theta_{im}^{\alpha\mu}\theta_{jn}^{\beta\nu}
        -(-1)^{\mu\nu+\alpha+\beta}
        \theta_{jn}^{\beta\nu}\theta_{im}^{\alpha\mu}\;,
        \label{pbraket}
        \ee
introduced in \cite{p10}. Note that Eq.~(\ref{e4}) is the statement of
the canonical quantization rule for the Green components $\theta$ on
the one hand, and the expression of the normal relative statistics
\cite{g-m,p10} on the other. The classical analogs of the self-adjoint
operators $\theta$ are obtained by setting $\hbar=0$ in Eq.~(\ref{e4}).

One also generalizes the definition of the parabracket to arbitrary
polynomials in $\theta$'s, according to:
        \be
        \pbra M,N \pket=MN-(-)^{\eta(M,N)}MN\;,
        \label{e5}
        \ee
where $M$ and $N$ are monomials:
        \bea
        M&:=&\theta_{i_1 m_1}^{\alpha_1\mu_1}\cdots\theta_{i_r m_r}^{
        \alpha_r \mu_r}\;,\nn\\
        N&:=&\theta_{j_1 n_1}^{\beta_1\nu_1}\cdots\theta_{j_s n_s}^{
        \beta_s \nu_s}\;,\nn\\
        \eta(M,N)&:=&(\sum_{k=1}^r\mu_k)(\sum_{l=1}^s\nu_l)+
                r\sum_{l=1}^s\beta_l+s\sum_{k=1}^r\alpha_k\;,
        \label{e6}
        \eea
and bilinearity of the parabracket. In the classical limit the parabracket
of any two polynomials vanishes identically.

In Ref.~\cite{p10}, it is also argued that in the Lagrangian formulation
of the para-classical mechanics, the Green components of the
$\pi b$ coordinate variables are $\theta^{\alpha \mu=0}_{i m=1}$. Thus,
one introduces a collective index $I =(i,m)$ which may take $(i=1,\cdots,
n_{\pi b};m=1)$ for $\mu=0$ and $(i=1,\cdots,n_{\pi f};m=1,2)$ for
$\mu=1$, and denote the Green components of the coordinate variables
by $\theta_I^{\alpha\mu}$.  The physical quantities, such as a Lagrangian,
is chosen from the algebra of polynomials in the coordinates
        \be
        \psi^\mu_I:=\sum_{\alpha=0}^1\theta^{\alpha\mu}_I
        \label{e7}
        \ee
and the velocities $\dot{\psi}^\alpha_I$. For computational convenience,
they are then expressed in terms of the Green components $\theta^{\alpha\mu
}_I$ and $\dot{\theta}^{\alpha\mu}_I$.

As a polynomial in (the classical) $\theta$'s and $\dot{\theta}$'s,
a Lagrangian must satisfy (up to total time derivatives) the following
conditions \cite{p10}:
        \begin{itemize}
        \item[1)] It must be real.
        \item[2)] It must be an even polynomial in both $\pi b$ ($\mu=0$) and
        $\pi f$ ($\mu=1)$ variables.
        \end{itemize}
To define the notion of reality in the algebra of polynomials in
$\theta$'s and $\dot\theta$'s (alternatively in $\psi$'s and  $\dot\psi$'s),
one first introduces a $*$--operation  satisfying:
        $$(\xi_{x_1}\cdots\xi_{x_n})^*=\xi_{x_n}\cdots\xi_{x_1}\;,$$
        $$(\lambda_1P_1+\lambda_2P_2)^*=\lambda_1^*P_1^*+\lambda_2^*P_2^*\;,$$
where $\xi_{x_i}$ are any of the generators: $\theta$'s and $\dot\theta$'s
(resp.\ $\psi$'s and $\dot\psi$'s), $\lambda_a\in\C$ with $a=1,2$,
$\lambda_a^*$ are their complex conjugates, and $P_a$ are polynomials in
$\xi_{x_i}$. Then a polynomial $P$ is defined to be real if $P^*=P$.

The classical dynamics of the system is given by the least action principle,
where the action functional has the form: $S=\int L dt$. This leads to the
analogs of the Euler-Lagrange equations:
        \be
        \frac{d}{dt}(L\frac{\ldel}{\partial \dot{\theta}})-
        L\frac{\ldel}{\partial\theta}=0\;.
        \label{e8}
        \ee
Here the indices on $\theta$'s are suppressed for simplicity and the
left partial derivatives with respet to $\theta$'s and $\dot\theta$'s
are defined in Refs.~\cite{o-k,p10}.

Having reviewed the basic elements of the Lagrangian formulation
of para-classical systems, we would like to address the question:
        \begin{itemize}
        \item[]
        {\em Does the algebraic structure of ($p=2$) parastatistical
        dynamical variables allow for a transformation of $\pi b$ variables
        into $\pi f$ variables and vice versa?}
        \end{itemize}

Unlike, the case of ordinary ($p=1$) fermi -- bose systems, where
the product of two fermi variables is a commutative algebraic object
and thus behaves as a bose variable, the algebraic structure of the
($p=2$) variables is too complicated to have such a simple grading.
Nevertheless, in view of the formalism developed in \cite{p10}, one can
easily respond to the above mentioned question in the positive.

To see this, consider the algebra $B$ of the real Green components generated
by $\xi_k^{\alpha\mu}$, and the algebra $A$ generated by:
        $$\gamma_k^\mu=\sum_{\alpha=0}^1 \xi_k^{\alpha\mu}\;.$$
The elements of $A$ (resp.\ $B$) will be used as non-dynamical parameters
added to the algebra of polynomials in dynamical variables
$\psi$'s and $\dot\psi$'s (resp.\ $\theta$'s and $\dot\theta$'s). Then
in the enlarged algebra, it is not difficult to check that the multiplication
of dynamical variables $\psi^\mu_I$ and $\dot\psi^\mu_I$ by the real
parameters:
        \be
        \gamma_k=\{\gamma^0_k,\gamma^1_k\}:=
        \sum_{\alpha,\beta=0}^1\{\xi^{\alpha \mu=0}_k,
        \xi^{\beta \mu=1}_k\}\;,
        \label{e9}
        \ee
changes their parity. Here there is no summation over the index $k$.
This can be easily verified by defining
$\psi^{\alpha\mu'}_{I}:=\psi^{\alpha\mu}_I \gamma_k$ and examining their
commutation properties by first decomposing them into their Green
components. One can further show that $\gamma_k$ commute with all the
parabose variables and anticommute with all the parafermi variables.

Presence of $\gamma_k$ allows for the existence of the $\pi SUSY$
transformations. We shall examine examples of such symmetry transformations
in the next section. We shall also introduce $\delta\gamma_k$ which are
analogs of the (fermionic) parameters of the infinitesimal supersymmetry
transformations.

\section{SUSY and $\pi$SUSY Oscillators}
A thorough discussion of the supersymmetric (SUSY) oscillator is offered
in Ref.~\cite{bd}. The Hamiltonian operator of one-dimensional SUSY
oscillator is the sum of the Hamiltonians of a fermi and a bose oscillators
with identical frequencies, i.e.,
        \bea
        \hat{H}&=&\hat{H}^0+\hat{H}^1\;,
        \label{e10}\\
        \hat{H}^0&:=&\frac{\omega}{2}\{\hat a^\dagger,\hat a\}\;,
        \label{e10.1}\\
        \hat H^1&:=&\frac{\omega}{2}[\hat \alpha^\dagger,\hat \alpha]\;.
        \label{e10.2}
        \eea
Here, $\hat a$ and $\hat a^\dagger$ (resp.\ $\hat\alpha$ and
$\hat\alpha^\dagger$) stand for the bosonic (resp.\ fermionic) annihilation
and creation operators and the hats are placed to distinguish the
quantum mechanical operators and the classical dynamical variables.

The combined system of two oscillators (\ref{e10}) serves as a simple
example of a supersymmetric system. To reveal the supersymmetry of
this system, we shall first switch to the self-adjoint operators:
        \bea
        \hat x&:=&\sqrt{\frac{\hbar}{2\omega}}\,(\hat a+\hat a^\dagger)\;,
        \nn\\
        \hat p&:=&-i\sqrt{\frac{\omega\hbar}{2}}\,(\hat a -\hat a^\dagger)\;,
        \label{e11}\\
        \hat
\psi_1&=&\sqrt{\frac{\hbar}{2}}\,(\hat\alpha+\hat\alpha^\dagger)\;,
        \nn\\
        \hat
\psi_2&=&-i\sqrt{\frac{\hbar}{2}}\,(\hat\alpha-\hat\alpha^\dagger)\;.\nn
        \eea
Then the Hamiltonian (\ref{e10}) takes the form:
        \be
        \hat H=\frac{1}{2}(\hat p^2+\omega^2 \hat x^2)+\frac{i\omega}{2}
        \epsilon_{mn}\hat \psi_m\hat \psi_n\;,
        \label{e12}
        \ee
where $\epsilon_{mn}$ are the components of the Levi Civita symbol.

The classical counterpart of the SUSY oscillator is obtained by dropping
the hats in the above relations and treating $x$ and $p$ as bosonic
(commuting or even) and $\psi_m$ as fermionic (anticommuting or
odd) supernumbers \cite{bd}, respectively.

The classical SUSY oscillator may also be described using the Lagrangian:
        \be
        L=\frac{1}{2}(\dot{x}^2-\omega^2x^2)+\frac{i}{2}\delta_{mn}
        (\psi_m\dot \psi_n-\dot
\psi_m\psi_n)-\frac{i\omega}{2}\epsilon_{mn}\psi_m
        \psi_n\;,
        \label{e13}
        \ee
where $m,n=1,2$. Then it is an easy exercise to check that this Lagrangian
is invariant (up to total time derivatives) under the transformation:
        \bea
        \delta x&=&i\psi_m\,\delta\zeta_m \;, \label{e14}\\
        \delta \psi_m&=&(\delta_{mn}\dot x+\omega\epsilon_{mn}
        x)\,\delta\zeta_n\;,
        \label{e15}
        \eea
where $\delta\zeta_n$ are ``infinitesimal'' fermionic supernumber parameters
\cite{bd}. The corresponding N\"other charges of this symmetry -- the
supercharges -- are given by:
        \be
        Q_m=\lambda (\delta_{mn}\dot x-\omega\epsilon_{mn}x)\psi_n\;,
        \label{charge}
        \ee
where $\lambda\in\C$ is an arbitrary non-zero coefficient. Upon quantization
of this system one can easily show that the supercharges, that generate
the supersymmetry transformations, and the Hamiltonian satisfy the
defining algebra of SQM. In particular, taking $\lambda=1/\sqrt{\hbar}$,
one has:
        \be
        \{ \hat Q_m,\hat Q_n\}=2\delta_{mn}\hat H\;.
        \label{sqm}
        \ee

Next, let us introduce the para-generalization of the SUSY oscillator.
We shall denote this by $\pi$SUSY oscillator for simplicity.

In general, the Hamiltonian for the parabose and parafermi oscillators
is given by Eqs.~(\ref{e10.1}) and (\ref{e10.2}), with $\hat a$ and
$\hat\alpha$, now, denoting the parabose and parafermi annihilation operators,
respectively, \cite{o-k}. Returning to our notation of Sec.~2, we set
$\hat a:=\hat a^{\mu=0}$ and $\hat\alpha:=\hat a^{\mu=1}$. In terms of the
self-adjoint operators: $\hat\psi^\mu:=\hat\psi^\mu_{i=1}$ of Eq.~(\ref{e7})
and their Green components $\theta_m^{\alpha\mu}:=\theta_{i=1,m}^{\alpha\mu}$,
we have:
        \bea
        \hat H^\mu&=&\frac{\omega}{2}[ (1-\mu)\delta_{mn}\hat\psi^\mu_m
        \hat\psi^\mu_n+i\mu\epsilon_{mn}\hat\psi_m^\mu\hat\psi_n^\mu ]\;,
        \label{e16}\\
        &=&\frac{\omega}{2}[ (1-\mu)\delta_{mn}
        \hat\theta_m^{\alpha\mu}\hat\theta_n^{\alpha\mu}+
        i\mu\epsilon_{mn}\hat\theta_m^{\alpha\mu}\hat\theta_n^{\alpha\mu}]\;,
        \label{e17}
        \eea
where $\mu=0,1$ correspond to $\pi b$ and $\pi f$ oscillators, respectively.

The ($p=2$) -- $\pi$SUSY oscillator is then defined by Eq.~(\ref{e10}):
        \be
        \hat H=\hat H^0+\hat H^1=\sum_{\alpha=0}^1\left\{
        [\frac{1}{2}(\hat\pi^\alpha)^2+\frac{\omega^2}{2}(\hat\chi^\alpha)^2]
        +\frac{i\omega}{2}\epsilon_{IJ}\,\hat\tau^\alpha_I\hat\tau^\alpha_J
        \right\}\;,
        \label{e18}
        \ee
where $\hat\chi^\alpha:=\hat\theta_1^{\alpha\mu=0}/\sqrt{\omega},~
\hat\pi^\alpha:=\sqrt{\omega}\hat\theta_2^{\alpha\mu=0}$
are the Green components of the $\pi b$ coordinate and momentum operators,
and $\hat\tau^\alpha_I:=\hat\theta^{\alpha\mu=1}_{m=I}$ are those of the
$\pi f$ coordinate operators.

The Lagrangian associated with the $\pi$SUSY oscillator is given by:
        \bea
        L&=&\frac{1}{2}(\dot x^2-\omega^2x^2)+\frac{i}{4}\delta_{IJ}
        (\psi_I\dot\psi_J-\dot\psi_I\psi_J)-\frac{i\omega}{2}
        \epsilon_{IJ}\,\psi_I\psi_J\;,
        \label{e20}\\
        &=&\sum_{\alpha=0}^1\left\{ \frac{1}{2}[(\dot \chi^\alpha)^2-
        \omega^2(\chi^\alpha)^2]+\frac{i}{2}\delta_{IJ}\tau^\alpha_I
        \dot\tau_J^\alpha-\frac{i\omega}{2}\epsilon_{IJ}\,\tau^\alpha_I
        \tau_J^\alpha\right\}\;,
        \label{e21}
        \eea
where $x=\sum_{\alpha=0}^1\chi^\alpha,~\psi_I=\sum_{\alpha=0}^1\tau_I^\alpha$
are the ($p=2$) $\pi b$ and $\pi f$ dynamical variables, respectively.

The form of the $\pi b$ and $\pi f$ kinetic terms in (\ref{e20}) is obtained
in Ref.~\cite{p10} in an attempt to consistently generalize the Peierls
bracket quantization scheme to the paraclassical systems.

The Peierls bracket quantization of this system leads to the following
paracommutation relations:
        \bea
        \pbra \hat\chi^\alpha,\hat\chi^\beta\pket&=&0\;,\nn\\
        \pbra\hat\chi^\alpha,\hat{\dot{\chi}}^\beta\pket&=&i\hbar
        \delta^{\alpha\beta}\;,\nn\\
        \pbra\hat{\dot\chi}^\alpha,\hat{\dot\chi}^\beta\pket&=&0\;,
        \label{e30}\\
        \pbra\hat\tau^\alpha_I,\hat\tau^\beta_J\pket&=&\hbar
        \delta^{\alpha\beta}\delta_{IJ}\;,\nn \\
        \pbra\hat\chi^\alpha,\hat\tau_I^\beta\pket&=&0\;,\nn \\
        \pbra \hat{\dot{\chi}}^\alpha,\hat\tau_I^\beta\pket&=&0\;,\nn
        \eea
which become identical with the canonical quantization relations (\ref{e4})
if one only considers the momenta $\pi^\alpha$ conjugate to $\chi^\alpha$ and
identifies them with $\dot{\chi}^\alpha$.

\section{Symmetries of the $\pi$SUSY Oscillator}

Setting $\tau^0_I=\tau^1_I$ and $\chi^0=\chi^1$ in
(\ref{e21}), one recovers the Lagrangian for the SUSY oscillator
(\ref{e13}). This may be used as a hint to seek similar symmetries
for the $\pi$SUSY oscillator.

Following this hint, consider the $\pi$-SUSY transformation:
        \bea
        \delta\chi^\alpha&=&i\tau^\alpha_J\,\delta\gamma_J\;,
        \label{e31}\\
        \delta\tau_I^\alpha&=&
        (\delta_{IJ}\,\dot\chi^\alpha+\omega
        \epsilon_{IJ}\,\chi^\alpha)\,\delta\gamma_J\;,
        \label{e32}
        \eea
where $\delta\gamma_J$ are the ``infinitesimal'' analogs of $\gamma_J$
of Eq.~(\ref{e9}). It is not difficult to check that the action
functional and therefore the dynamical equations remain invariant
under this transformation. Indeed, one finds:
        \be
        \delta L \propto \frac{d}{dt}(
        \dot\chi^\alpha\tau^\alpha_J -
        \omega\epsilon_{JI}\tau_I^\alpha\chi^\alpha)\,\delta\gamma_J\;.
        \label{e33}
        \ee
Thus the corresponding conserved charges have the form:
        \be
        Q^1_J=\lambda(
        \dot\chi^\alpha\tau^\alpha_J - \omega\epsilon_{JI}\tau_I^\alpha
        \chi^\alpha)\;.
        \label{e34}
        \ee
Here the superscript ``1'' is placed for later use and $\lambda$ is a
non-zero numerical coefficient.

In the remainder of this paper, we shall set $\hbar=1$ for simplicity.

The quantum analog of $Q^1_J$ with an appropriate normalization is given
by:
        \be
        \hat Q^1_J:=\hat{\dot\chi}^\alpha
        \hat\tau^\alpha-\omega\epsilon_{JI}\hat\chi^\alpha
        \hat\tau_I^\alpha\;.
        \label{e35}
        \ee
In view of the paracommutation relations (\ref{e30}), it is not difficult
to check that $Q_J^1$ generate the transformations (\ref{e31}) and (\ref{e32}),
i.e.,
        \bea
        \pbra \hat\chi^\alpha,\hat Q_J^1\delta\gamma_J\pket&=&
        i\hat\tau^\alpha_J\delta\gamma_J\:=\: \delta\hat\chi^\alpha\;,
        \label{e36}\\
        \pbra \hat\tau^\alpha_I,\hat Q_J^1\delta\gamma_J\pket&=&
        (\delta_{IJ}\hat{\dot{\chi}}^\alpha+\omega
        \epsilon_{IJ}\hat\chi^\alpha)\,\delta\gamma_J\:=\:
        \delta\hat\tau^\alpha_I\;,
        \label{e37}
        \eea
and that they satisfy the defining algebra of SQM, namely:
        \be
        \{ \hat Q_I^1,\hat Q_J^1\}=2\delta_{IJ}\hat H\;.
        \label{e38}
        \ee
Note also that $\hat Q^1_I$ are self-adjoint operators by construction
(\ref{e35}).

Another important point in handling ($p=2$) para-dynamical systems is
that the Green components are not the physical dynamical variables.
In other words, one must be able to express all physical quantities in terms
of the variables, $x,~\dot x,~\psi_I,$ and $\dot\psi_I$. This also
applies to the $Q_I^1$. In fact, one can show that:
        \be
        \hat Q_J^1=\frac{1}{2}\{
        \delta_{JK}\hat{\dot x}-\omega\epsilon_{JK}\hat x\:,\:\hat\psi_K\}\;.
        \label{e39}
        \ee
Here use is made of the identities:
        \be
        \hat{\dot\chi}^\alpha \hat\tau^\alpha_I=\frac{1}{2}\{\hat{\dot x},
        \hat\psi\}\;,~~~~~~
        \hat\chi^\alpha \hat\tau^\alpha_I=\frac{1}{2}
        \{\hat x,\hat\psi_I\}\;.
        \label{e40}
        \ee

The $\pi$SUSY transformations (\ref{e31}) and (\ref{e32}) mix the Green
components $\chi^\alpha$ and $\tau_I^\alpha$ with the same Green index
$\alpha$. Since the Green components are not physical quantities, there
must be no difference between say $\tau^0_I$ and $\tau^1_I$. This suggests
the possibility of symmetry transformations which mix $\chi^\alpha$ with
$\tau^{\alpha+1}$. Here the values of the Green indices is taken in $\Z_2$,
i.e., they are calculated modulo $2$. The following is such a symmetry
transformation:
        \bea
        \delta\chi^\alpha&=&-i\tau^{\alpha+1}_J\,\delta\gamma_J\;,
        \label{e41}\\
        \delta\tau_I^\alpha&=&
        (\delta_{IJ}\dot\chi^{\alpha+1}+\omega
        \epsilon_{IJ}\chi^{\alpha+1})\,\delta\gamma_J\;.
        \label{e42}
        \eea

The associated conserved charges to this symmetry are given by
        \be
        Q^2_J=\lambda'(
        \tau^\alpha_J\dot\chi^{\alpha+1}-
        \omega\epsilon_{JK}\tau_K^\alpha\chi^{\alpha+1})\;,
        \label{e43}
        \ee
where the summation over $\alpha$ is understood. Quantizing the system and
taking:
        \be
        \hat Q^2_J:=i(\hat\tau^\alpha_J\hat{\dot{\chi}}^{\alpha+1}-
        \omega\epsilon_{JK}\hat\tau_K^\alpha\hat\chi^{\alpha+1})\;,
        \label{e44}
        \ee
one obtains another set of self-adjoint $\pi$SUSY charges.
They generate the transformations (\ref{e41}) and (\ref{e42}) and are
expressed in terms of the physical variables $x$ and $\psi$ according to:
        \be
        \hat Q^2_J=
        \frac{-i}{2}[\delta_{JK}\hat{\dot x}-\omega\epsilon_{JK}\hat x\:,\:
        \hat\psi_K]\;.
        \label{e45}
        \ee
Here use is made of the identities:
        \be
        \hat\tau_I^\alpha\hat{\dot\chi}^{\alpha+1}=\frac{1}{2}
        [\hat\psi_I,\hat{\dot x}]\;,~~~~~~
        \hat\tau_I^\alpha\hat\chi^{\alpha+1}=\frac{1}{2}[\hat\psi_I,\hat x]\;.
        \label{e46}
        \ee
Furthermore, the superalgebra relation:
        \be
        \{ \hat Q^2_I,\hat Q^2_J\}=2\delta_{IJ}\hat H\;,
        \label{e47}
        \ee
also holds.

The next natural step in the study of the symmetries of the $\pi$SUSY
oscillator is to investigate the algebraic properties of both types
of $\pi$SUSY's. Proceeding in this direction, one finds:
        \be
        \pbra \hat Q_J^a,\hat Q_K^b\pket=
        \{ \hat Q_J^a,\hat Q_K^b\}=2\delta_{JK}\delta^{ab}\hat H
                -2\epsilon^{ab}\epsilon_{JK}\hat {\cal Q}\;,
        \label{e48}
        \ee
where
        \be
        \hat{\cal Q}:=i\omega\hat\chi^\alpha\hat{\dot\chi}^{\alpha+1}+
        \frac{\omega}{2}\hat\tau_I^\alpha\hat\tau_I^{\alpha+1}
        \label{e49}
        \ee
is another (self-adjoint) conserved charge.

Repeating this procedure, i.e., including ${\cal Q}$ in the set
of the generators of symmetries and investigating the parabracket
of ${\cal Q}$ and other generators, one obtaines
        \be
        \pbra \hat Q^a_J,\hat{\cal Q}\pket=[ \hat Q^a_J,\hat{\cal Q}]=0\;.
        \label{e50}
        \ee
Thus the {\em superalgebra} consisting of the generators of $\pi$SUSY
of the $\pi$SUSY oscillator closes. Summarizing the superalgebra
relations, one has:
        \bea
        \pbra \hat Q_J^a,\hat H\pket&=&[\hat Q_J^a,\hat H]\:=\:0\nn\\
        \pbra \hat Q_J^a,\hat Q_K^b\pket&=&
        \{ \hat Q_J^a,\hat Q_K^b\}\: =\: 2\delta_{JK}\delta^{ab}\hat H
                -2\epsilon^{ab}\epsilon_{JK}\hat {\cal Q}\;,\nn\\
        \pbra \hat{\cal Q},\hat H\pket&=&[ \hat{\cal Q},\hat H]\:=\: 0
        \label{e60}\\
        \pbra \hat Q^a_J,\hat{\cal Q}\pket&=&[ \hat Q^a_J,\hat{\cal Q}]
        \:=\: 0\;.\nn
        \eea
The generators $Q_J^a$ behave as the ``odd'' elements of the super
Lie algebra and $H$ and ${\cal Q}$ as the ``even'' (central) elements.

The (central) charge ${\cal Q}$ is also expressed in terms of the physical
variables. One has:
        $$ {\cal Q}=\frac{i\omega}{2}(\hat x\hat{\dot x}-\hat{\dot x}\hat x)
        +\frac{\omega}{2}\delta_{IJ}\hat\psi_I\hat\psi_J\;. $$
Here, one uses the following identities:
        $$\delta_{IJ}\hat\psi_I\hat\psi_J=
        \hat\tau^\alpha_I\hat\tau^{\alpha+1}_I+2\;,~~~~~~
        \hat x\hat{\dot x}-\hat{\dot x}\hat x=
        2\hat\chi^\alpha\hat{\dot\chi}^{\alpha+1}+2i\;.$$

One can also examine the symmetry transformations generated by ${\cal Q}$.
These are obtained by computing:
        \bea
        \pbra \hat\chi^\alpha,{\cal Q}\delta\epsilon\pket&=&
        \{\hat\chi^\alpha,{\cal Q}\} \delta\epsilon\:=\:\omega
        \hat\chi^{\alpha+1}\delta\epsilon\;,\nn\\
        \pbra \hat\tau^\alpha_I,{\cal Q}\delta\epsilon\pket&=&
        \{\hat\tau^\alpha_I,{\cal Q}\}\delta\epsilon\:=\:\omega
        \hat\tau^{\alpha+1}_I\delta\epsilon\;.\nn
        \eea
Thus:
        $$\delta_{\cal Q}\chi^\alpha=\omega\,
        \chi^{\alpha+1}\delta\epsilon\;,~~~~~~
        \delta_{\cal Q}\tau^\alpha_I=\omega\,
        \tau^{\alpha+1}_I\delta\epsilon\;.$$
Here $\delta\epsilon$ is an infinitesimal commuting parameter. In terms
of the physical dynamical variables, one has:
        $$\delta_{\cal Q}x=\omega\,x\,\delta\epsilon\,\;,~~~~~~
        \delta_{\cal Q}\psi_I= \omega\,\psi_I\,\delta\epsilon\,\;.$$

We would like to conclude this section by emphasizing the enormous
advantage of using parabracket (\ref{pbraket}) in performing the
tedius computations necessary for establishing the superalgebra relations
Eqs.~(\ref{e60}). The details of these computations have been omitted
due to the space limitations.

\section{Degeneracy Structure of General SPQM}
Let us first define SPQM:
        \begin{itemize}
        \item[] {\bf Definition:} {\em Let ${\cal H}$ be a $\Z_2$-graded
        Hilbert space with grading involution {\large $\hat\tau$}. Then
        a quantum mechanical system with ${\cal H}$ as the Hilbert space
        and self-adjoint symmetry generators $\hat Q_{I_n}^{a_n}$,
        $\hat{\cal Q}_n$, $n=1,\cdots N$, ~ $I_n,a_n=1,2$,
        and the Hamiltonian operator $\hat H$ satisfying the super Lie algebra
        relations:
                \bea
                [\hat Q_{J_n}^{a_n},\hat H]&=&[\hat{\cal Q}_n,H]\:=\:
                [ \hat Q^{a_n}_{J_n},\hat{\cal Q}]\:=\:0\label{e70}\\
                \{ \hat Q_{J_n}^{a_n},\hat Q_{K_m}^{b_m}\}&=&\delta_{nm}(
                2\delta_{J_nK_n}\delta^{a_nb_n}\hat H
                -2\epsilon^{a_nb_n}\epsilon_{J_nK_n}\hat{\cal Q}_n)\;,
                \label{e71}
                \eea
        and parity properties:
                \be
                \{\mbox{\large$\hat\tau$},\hat Q_{I_n}^{a_n}\}=0\;,~~~~~
                [\mbox{\large$\hat\tau$},\hat{\cal Q}_n]=
                [\mbox{\large$\hat\tau$},\hat H]=0\;,
                \label{e72}
                \ee
        for all $I_n,a_n$ and $n=1,\cdots,N$, is called a ($p=2$) --
        supersymmetric paraquantum mechanical (SPQM) system of type $N$.}
        \end{itemize}
In this section, we shall present a detailed analysis of the spectrum
degeneracy structure of general ($p=1$)-SPQM systems of type $N=1$.

For $N=1$ we suppress the index $n=1$ and recover the super Lie algebra
of the $\pi$SUSY oscillator, i.e., Eqs.\ (\ref{e60}). For simlicity we
shall drop the hats and introduce the notation:
        $$Q_1\equiv Q_1^1\;,~~~Q_2\equiv Q_2^1\;,~~~Q_3\equiv Q_1^2\;,
        ~~~Q_4\equiv Q_2^2\;.$$
Then Eqs.~(\ref{e60}) are written as:
        \bea
        Q_i^2&=&H\;,\label{e81}\\
        \{ Q_1,Q_2\}&=&0\;,\label{e82}\\
        \{ Q_1,Q_3\}&=&0\;,\label{e83}\\
        \{ Q_1,Q_4\}&=&-2 {\cal Q}\;,\label{e85}\\
        \{ Q_2,Q_3\}&=&2 {\cal Q} \;,\label{e84}\\
        \{ Q_2,Q_4\}&=&0\;,\label{e86}\\
        \{ Q_3,Q_4\}&=&0\;,\label{e87}\\
        {[} Q_i,{\cal Q}{]}&=&0\;,\label{e88}
        \eea
where $i=1,2,3,4$.

Next, we use the simultaneous eigenstate vectors $|E,q_1,q\kt$, with
$E,q_1,q\in\R$, of $H,~ Q_1$ and ${\cal Q}$ to span the Hilbert space.
We shall assume that these state vectors form an orthonormal basis and
attempt to represent all the relevant operators in this basis. These
properties are summarized by the following set of relations:
        \bea
        H|E,q_1,q\kt&=&E|E,q_1,q\kt\;,~~~~~~
        Q_1|E,q_1,q\kt\:=\: q_1|E,q_1,q\kt\;,
        \label{e91}\\
        {\cal Q}|E,q_1,q\kt&=&q|E,q_1,q\kt\;,~~~~~~
        \br E',q'_1,q'|E,q_1,q\kt\:=\:
        \delta_{E'E}\delta_{q'_1,q_1}\delta_{q'q}\;.
        \label{e92}
        \eea

A simple consequence of Eq.~(\ref{e81}) with $i=1$, is that the energy
spectrum is non-negative. Furthermore, for any energy level $E$, one has:
        \bea
        q_1&=&\pm\sqrt{E}\;,
        \label{e101}\\
        |q_1,q\kt=0&\Leftrightarrow&|-q_1,q\kt=0\;,
        \label{e102}\\
        Q_2|q_1,q\kt&=&C_2(q_1,q)|-q_1,q\kt\;,~~~~~~C_2(q_1,q)\in\C-\{0\}\;,
        \label{e109}\\
        Q_3|q_1,q\kt&=&C_3(q_1,q)|-q_1,q\kt\;,~~~~~~C_3(q_1,q)\in\C-\{0\}\;,
        \label{e110}
        \eea
where use is made of Eqs.~(\ref{e81}) -- (\ref{e83}) and abbreviation
$|q_1,q\kt$ is used for $|E,q_1,q\kt$. Enforcing Eq.~(\ref{e84}), one finds:
        \be
        C_2(q_1,q)C_3(-q_1,q)+C_3(q_1,q)C_2(-q_1,q)=2 q\;.
        \label{e111}
        \ee
Then by acting both sides of Eqs.~(\ref{e81}), with $i=2,3$, on $|q_1,q\kt$,
one has:
        \be
        C_2(q_1,q)C_2(-q_1,q)=E\;,~~~~~~C_3(q_1,q)C_3(-q_1,q)=E\;.
        \label{e115}
        \ee
Next, we calculate:
        $$E=(\br q_1,q|Q_2)(Q_2|q_1,q\kt)=C_2(q_1,q)^*C_2(q_1,q)\;.$$
A similar relation holds for $C_3$. These relations together with
Eqs.~(\ref{e115}) imply:
        \be
        C_2(\pm q_1,q)=\sqrt{E}\,e^{\pm i\alpha_2(q)}\;,~~~~~~
        C_3(\pm q_1,q)\:=\:\sqrt{E}\,e^{\pm i\alpha_3(q)}\;.
        \label{e116}
        \ee

Combining the latter equations with Eq.~(\ref{e111}), one is led to:
        \be
        \frac{C_2(q_1,q)}{C_3(q_1,q)}+\frac{C_3(q_1,q)}{C_2(q_1,q)}
        =\frac{2q}{E}\;.
        \label{e117}
        \ee
Eqs.~(\ref{e116}) and (\ref{e117}), in turn, yield:
        \be
        \cos[\alpha_2(q)-\alpha_3(q)]=\frac{q}{E}\;.
        \label{e121}
        \ee

Next, we act both sides of Eqs.~(\ref{e85}), (\ref{e86}), and (\ref{e87})
on $|q_1,q\kt$ on the left. This gives rise to:
        \bea
        Q_1Q_4|q_1,q\kt&=&-q_1Q_4|q_1,q\kt-2 q|q_1,q\kt\;,
        \label{e112}\\
        Q_2Q_4|q_1,q\kt&=&-\sqrt{E}\,e^{i\alpha_2}Q_4|-q_1,q\kt\;,
        \label{e119}\\
        Q_3Q_4|q_1,q\kt&=&-\sqrt{E}\,e^{i\alpha_3}Q_4|-q_1,q\kt\;.
        \label{e120}
        \eea
To pursue our analysis further, we express the action of $Q_4$ on the basic
kets $|q_1,q\kt$ as the following linear combination:
        \be
        Q_4|q_1,q\kt=: a(q_1,q)|\sqrt{E},q\kt+b(q_1,q)|-\sqrt{E},q\kt
        \;.
        \label{e130}
        \ee
where $a$ and $b$ are complex numbers {\em a priori} depending on $q_1$,
$q$ and of course $E$. Substituting this expression in Eq.~(\ref{e112}),
one finds:
        \be
        a(q_1=\sqrt{E},q)=-\frac{q}{\sqrt{E}}\,~~~~~
        a(q_1=-\sqrt{E},q)=0\;.
        \label{e131}
        \ee
Repeating the same procedure for Eqs.~(\ref{e119}) and (\ref{e120}),
and performing the simple algebra, one finally obtains:
        \bea
        b(q_1=\sqrt{E},q)&=&0\;,~~~~~b(q_1=-\sqrt{E},q)\:=\:-a(q_1=
        \sqrt{E},q)=\frac{q}{\sqrt{E}}\;,
        \label{e6.1}\\
        e^{i\alpha_2(q)}\:=\:\pm 1 &=&e^{i\alpha_3(q)}\;.
        \label{e6.2}
        \eea
The last pair of equations together with Eq.~(\ref{e121}) imply:
        \be
        q=E\eta\;,~~~~~~~~\eta=\pm 1\;.
        \label{eq}
        \ee

Having obtained all the unknowns of our construction and appealing to the
gauge freedom of the phases of the initial basic eigenstate vectors -- which
allows us to set, say, $\alpha_2=0$ so that $e^{i\alpha_3}=\eta$ --
we are in a position to present matrix reperesentations of all the
charges. However, before presenting these representations, we would
like to remark that although $|\sqrt{E},q\kt\neq 0\Leftrightarrow
|-\sqrt{E},q\kt\neq 0$, this relation does not imply that
$|\pm\sqrt{E},q\kt\neq 0$ for some $q$, i.e., in general it may be the
case that for some values of $E$ the state vectors corresponding to
either $q=+E$ or $q=-E$ vanish. In this case $E$ will be doubly
degenerate. Otherwise it will be quadruply degenerate. For the
latter case the symmetry generators are represented by:
        \bea
        \left. Q_1\right|_{{\cal H}_E}&=&\sqrt{E}\left(\begin{array}{cccc}
                1  &  0  &     &    \\
                0  & -1  &     &    \\
                   &     &  1  &  0 \\
                   &     &  0  &  -1
                \end{array}\right)\nn\\
        \left. Q_2\right|_{{\cal H}_E}&=&
                \sqrt{E}\left(\begin{array}{cccc}
                0  &  1  &     &    \\
                1  &  0  &     &    \\
                   &     &  0  &  1 \\
                   &     &  1  &  0
                \end{array}\right)\nn\\
        \left. Q_3\right|_{{\cal H}_E}&=&
                \sqrt{E}\left(\begin{array}{cccc}
                0  &  1  &     &    \\
                1  &  0  &     &    \\
                   &     &  0  & -1 \\
                   &     &  -1 &  0
                \end{array}\right)\nn\\
        \left. Q_4 \right|_{{\cal H}_E}&=&
                \sqrt{E}\left(\begin{array}{cccc}
                -1  &  0  &     &    \\
                 0  &  1  &     &    \\
                    &     &  1  &  0 \\
                    &     &  0  &  -1
                \end{array}\right)\nn\\
        \left. {\cal Q}\right|_{{\cal H}_E}&=&
                E\left(\begin{array}{cccc}
                1  &  0  &     &    \\
                0  &  1  &     &    \\
                   &     &  -1 &  0 \\
                   &     &  0  & -1
                \end{array}\right)\nn\\
        \left. H\right|_{{\cal H}_E}&=&
                E\left(\begin{array}{cccc}
                1  &  0  &     &    \\
                0  &  1  &     &    \\
                   &     &  1  &  0 \\
                   &     &  0  &  1
                \end{array}\right)\;.\nn
        \eea
Here we have identified:
        \bea
        |\sqrt{E},E\kt&=&\left(\begin{array}{c}
        1\\0\\0\\0
        \end{array}\right)\;,~~~~~~
        |-\sqrt{E},E\kt\:=\:\left(\begin{array}{c}
        0\\1\\0\\0
        \end{array}\right)\;,\nn\\
        |\sqrt{E},-E\kt&=&\left(\begin{array}{c}
        0\\0\\1\\0
        \end{array}\right)\;,~~~~~~
        |-\sqrt{E},-E\kt\:=\:\left(\begin{array}{c}
        0\\0\\0\\1
        \end{array}\right)\;,\nn
        \eea
the empty blocks consist of vanishing entries, and ${\cal H}_E$ denotes
the degeneracy Hilbert space associated with the energy $E>0$.

In view of Eqs.~(\ref{e72}), we can also write down the matrix representation
of the involution (chirality) operator in this basis. The result is given
by
        $$
        \left. \mbox{\large$\tau$}\right|_{{\cal H}_E}=
                \left(\begin{array}{cccc}
                0            &  -i\epsilon_1  &             &    \\
                i\epsilon_1  &      0         &             &    \\
                             &                &  0          &-i\epsilon_2\\
                             &                &i\epsilon_2  &   0
                \end{array}\right)\;,
        $$
where $\epsilon_1,\epsilon_2=\pm 1$.

It is an easy exercise to diagonalize the chirality involution(s) and to
find out that in the diagonal form it has the form:
        \be
        \left. \mbox{\large$\tau$}\right|_{{\cal H}_E}=
        {\rm diag}(1,-1,1,-1)\;.
        \label{e150}
        \ee
This implies that the quadruply degenerate (positive) energy levels
involve two odd (parafermionic) and two even (parabosonic) state vectors.

The representations of the symmetry generators and the involution
operator for the doubly degenerate energy levels are given by either
of the upper-left or lower-right blocks in the above list of matrix
representations, according to whether $|\pm\sqrt{E},-E\kt=0$ or
$|\pm\sqrt{E},+E\kt=0$, respectively. The situation is analogous to
the ordinary supersymmetric case, \cite{p8}.

The following lemma summarizes our results concerning ($p=2$)-SPQM:
        \begin{itemize}
        \item[] {\bf Lemma 1:} {\em
        The energy spectrum of any ($p=2$) supersymmetric paraquantum system
        is non-negative. The zero-energy eigenvalue, if exists, is
        non-degenerate\footnote{This is true provided that other quantum
        numbers are not present.}. The positive energy levels are
        either doubly or quadruply degenerate. They consist of pairs of
        odd and even parity eigenstates.}
        \end{itemize}
Moreover, one can define the Witten index according to
        $${\rm index_{Witten}}:={\rm trace}(\mbox{\large$\tau$})\;,$$
and use Eq.~(\ref{e150}) to conclude that it counts the difference
of the number of even and odd zero-energy states, and that it is a
topological invariant.

\section{Hilbert Space Structure of the $\pi$SUSY Oscillator}
Ref.~\cite{sud} offers an analysis of the energy eigenstates
of the one-dimensional parabose oscillator of arbitrary order $p$.
In the following, we shall use the results of \cite{sud}, with $p=1$,
to construct a complete set of eigenstate vectors for the
$\pi$SUSY oscillator.

The Hilbert space of the one-dimensional ($p=1$) $\pi$b oscillator
is constructed using the following set of orthonormal energy eigenstate
vectors:
        \be
        |n\kt:=\frac{1}{\sqrt{2^n[\frac{n}{2}]![\frac{n+1}{2}]!}}
               \: a^{\dagger n}\,|0\kt\;,
        \label{e200}
        \ee
where $a^\dagger$ and $|0\kt$ are the $\pi$b creation operator and
the vacuum (ground) state, respectively, and $[k]$ stands for the
largest integer smaller than or equal to $k\in\R$. One also has:
        \bea
        H^0|n\kt&=&(n+1)\omega\;,
        \label{e201}\\
        a|n\kt&=&\sqrt{2[\frac{n}{2}]+1}\:|n-1\kt\;,
        \label{e202}\\
        a^\dagger|n\kt&=&\sqrt{2[\frac{n}{2}]+2}\:|n+1\kt\;.
        \label{e203}
        \eea

For the $\pi$SUSY oscillator, one has also the $\pi$f creation and
annihilation operators. These have the property that $a^3=0$. So there
is an apparent triple grading intrinsic to the ($p=2$) $\pi$f operators.
This has been used quite often in the context of parasupersymmetry.
In the following we shall demonstrate that this is not the case for
the $\pi$SUSY oscillator as one might expect in view of the treatment
of Sec.~6.

It turns out that the following energy eigenstates form an orthonormal
basis for the Hilbert space:
        \be
        \begin{array}{cc}
        |n,1\kt:=
        \frac{1}{\sqrt{2^n[\frac{n}{2}]![\frac{n+1}{2}]!}}\:
        a^{\dagger n}|0\kt\;, ~~~~~~
        &(n\geq 0)\\
        |n,2\kt:=
        \frac{1}{\sqrt{2^n[\frac{n}{2}]![\frac{n-1}{2}]!}}
        \:\alpha^\dagger a^{\dagger n-1}|0\kt\;, ~~~~
        &(n\geq 1) \\
        |n,3\kt:=
        \frac{1}{\sqrt{2^n[\frac{n-1}{2}]![\frac{n-2}{2}]!}}
        \:\alpha^{\dagger 2}a^{\dagger n-2}|0\kt\;, ~
        &(n\geq 2) \\
        |n,4\kt:=
        \frac{1}{\sqrt{2^n[\frac{n}{2}]![\frac{n-1}{2}]!}}\:
        a^\dagger\alpha^\dagger a^{\dagger n-2}|0\kt\;
        &(n\geq 2)
        \end{array}
        \label{e204}
        \ee
Note that the state vector $|n,1\kt$ is the same as $|n\kt$ of
Eq~(\ref{e200}). To establish the orthonormality of $\{ |n,a\kt~:~
a=1,2,3,4\}$, one needs to use the following set of paracommutation
relations:
        \bea
        \alpha\,\alpha^\dagger\,a^\dagger&=&-a^\dagger \,\alpha^\dagger\,
        \alpha+2a^\dagger\;,
        \label{e210}\\
        \alpha\,\alpha^{\dagger 2}&=&-\alpha^{\dagger 2}\,
        \alpha+2\alpha^\dagger\;,
        \label{q2.4}\\
        a\,a^\dagger\,\alpha^\dagger&=&\alpha^\dagger\,a^\dagger\,a+
        2\alpha^\dagger\;,
        \label{e212}\\
        \alpha\, a\, a^\dagger&=&a^\dagger\, a\,\alpha+2\alpha\;,
        \label{e213}\\
        \alpha\,a \, \alpha^\dagger&=&-\alpha^\dagger\, a\,\alpha\;,
        \label{q23.1}
        \eea
and the identity:
        \be
        \alpha\, a^{\dagger n}|0\kt=0\;.
        \label{e215}
        \ee
Relations (\ref{e210})--(\ref{e215}) are most easily proved in the Green
representation.

Furthermore, it is not difficult to check that indeed $|n,a\kt$ are
energy eigenvectors, i.e.,
        \be
        H|n,a\kt=E_n|n,a\kt\;,~~~~~{\rm with}~~~~~E_n:=n\omega\;.
        \label{e220}
        \ee
Finally, it is possible to show that $|n,a\kt$ form a complete set of
state vectors. This involves some lengthy algebraic manipulations.
The completeness of $\{ |n,a\kt\}$ results from the following set
of relations:
        \bea
        a\,|n,1\kt&=&\sqrt{2[\frac{n}{2}]+1}\:|n-1,1\kt\;,\nn\\
        a^\dagger\,|n,1\kt&=&\sqrt{2[\frac{n}{2}]+2}\:|n+1,1\kt\;,\nn\\
        \alpha\,|n,1\kt&=&0\nn\\
        \alpha^\dagger\,|n,1\kt&=&\sqrt{2}\:|n+1,2\kt\;,\nn\\
        a\,|n,2\kt&=&\sqrt{\frac{[(n-1)/2](2[n/2]-1)}{[n/2]}}\:|n-1,5\kt\;,
        \nn\\
        a^\dagger\,|n,2\kt&=&\sqrt{2[(n+1)/2]}\: |n+1,5\kt\;,\nn\\
        \alpha\,|n,2\kt&=&\sqrt{2}\,|n-1,1\kt\;,\nn\\
        \alpha^\dagger\,|n,2\kt&=&\sqrt{2}\,|n+1,3\kt\;,\nn\\
        a\,|n,3\kt&=&-\sqrt{2[n/2]}\:|n-1,3\kt\;,\nn\\
        a^\dagger\,|n,3\kt&=&-\sqrt{2[n/2]}\,|n+1,3\kt\;,\nn\\
        \alpha\,|n,3\kt&=&\sqrt{2}\,|n-1,2\kt\;,\nn\\
        \alpha^\dagger\,|n,3\kt&=&0\;,\nn\\
        a\,|n,4\kt&=&
        \sqrt{\frac{[(n-1)/2](2[n/2]-1)+2}{[n/2]}}\:|n-1,2\kt\;,\nn\\
        a^\dagger\,|n,4\kt&=&\sqrt{2[(n+1)/2]}\,|n+1,2\kt\;,\nn\\
        \alpha\,|n,4\kt&=&0\;,\nn\\
        \alpha^\dagger\,|n,4\kt&=&0\;,\nn
        \eea
where in addition to Eqs.~(\ref{e202})--(\ref{q23.1}),
the paracommutation relations:
        \bea
        a\,\alpha^\dagger\,a^\dagger&=&a^\dagger\,\alpha^\dagger\,a\;,\nn\\
        a\,\alpha^{\dagger 2}&=&-\alpha^{\dagger 2}\,a\;,\nn\\
        \alpha\, a^\dagger\,\alpha^\dagger&=&-\alpha^\dagger\,a^\dagger\,
        \alpha\;,\nn\\
        \alpha^\dagger\, a^\dagger\,\alpha^\dagger&=&0\;,\nn
        \eea
are also used.

To demonstrate the method of proof of such relations using the
Green representation, a proof of the last equation is offered
in the following. First note that
        $$ a=\sum_{\alpha=0}^1\zeta^{\alpha 0}\;,~~~
        \alpha=\sum_{\beta=0}^1\zeta^{\beta 1}\;,$$
        \be
        \pbra \zeta^{\alpha \mu},\zeta^{\beta\nu}\pket=0\;,~~~
           \pbra \zeta^{\alpha \mu},\zeta^{\beta\nu\dagger}\pket=
           \delta^{\alpha\beta}\delta^{\mu\nu}\;.
        \label{e301}
        \ee
Here use is made of Eqs.~(\ref{e1})--(\ref{e4}). Next, one has:
        \bea
        \alpha^\dagger\, a^\dagger\,\alpha^\dagger&=&
        \sum_{\alpha,\beta,\gamma}
        \zeta^{\alpha 1\dagger}\zeta^{\beta 0\dagger}\zeta^{\gamma 1 \dagger}
        \nn\\
        &=&\sum_{\alpha,\beta,\gamma}(-1)^{1+\beta+\gamma}
        \zeta^{\beta 0 \dagger}\zeta^{\gamma 1\dagger}\zeta^{\alpha 1
        \dagger}\nn\\
        &=&-\sum_{\alpha,\beta,\gamma}
        \zeta^{\gamma 1\dagger}\zeta^{\beta 0\dagger}\zeta^{\alpha 1\dagger}
        \nn\\
        &=&-\alpha^\dagger\, a^\dagger\,\alpha^\dagger\:=\:0\;,
        \eea
where in the second and third equalities use is made of
Eqs.~(\ref{pbraket}) and (\ref{e301}).

This concludes our investigation of the energy eigenstates of the
$\pi$SUSY oscillator. We summarize the results of this section in the form
of the following lemma:
        \begin{itemize}
        \item[] {\bf Lemma~2:} {\em
        The energy spectrum of the $\pi$SUSY oscillator consists of
        a zero energy non-degenerate ground state (represented by
        $|n=0,1\kt$), a doubly degenerate first excited state (level) of
        energy $E_1=\omega$ (with state vectors $|n=1,1\kt$ and
        $|n=1,2\kt$), and higher excited states of $E_n=n\omega$
        ($n\geq 2$) which are quadruply degenerate (with state vectors
        $|n,a\kt\;,a=1,2,3,4$.)}
        \end{itemize}
This confirms our general results of Sec.~6.

\section{Conclusion}
There are dynamical systems involving ($p=2$) parastatistical degrees
of freedom and symmetry transformations which mix the parabose and parafermi
dynamical variables. The mixing which signifies a parabose -- parafermi
supersymmetry is shown to be present because of the non-trivial algebraic
properties of such variables.

Having established the meaningfullness of the parabose -- parafermi
supersymmetry ($\pi$SUSY), one can investigate its relation with the ordinary
(bose -- fermi) supersymmetry and the parasupersymmetry. The simple example
of an oscillator consisiting of a parabosonic and a parafermionic sector is
used to demonstrate the nature of $\pi$SUSY. This oscillator possesses
two ordinary supersymmetries. The study of the combined set of generators
of these supersymmetries leads to the introduction of a central charge.
Thus, it seems that there is no direct relation between parabose -- parafermi
supersymmetry and parasupersymmetry.

The oscillator considered in this article also serves as a useful example
to demonstrate the practical importance of the parabracket introduced
in \cite{p10}. Moreover, it is remarkable to check that indeed all the
conserved charges depend on the physical dynamical variables and not on their
Green components. This is quite non-trivial, for all the calculations
are performed using the Green components. In view of these observations,
one may conclude that there is no anomalous phenomena stemming from
the unusual parastatistical nature of the ($p=2$) dynamical variables.
In fact, it is shown that for example the $\pi$SUSY oscillator has a larger
symmetry than the ordinary SUSY oscillator.

Another interesting observation regarding the symmetries of the
$\pi$SUSY oscillator is that {\em a priori} there is
no parity associated with the quantities (polynomials) constructed
out of the parastatistical variables, nevertheless the conserved charges
and the Hamiltonian do possess parities, and they do form a super Lie
algebra. This may be seen as the primary reason why one does not need
trilinear algebraic relations between the symmetry generators.
The latter has been shown \cite{r-s} to be unavoidable for an oscillator
consisting of ordinary bosons and ($p=2$) parafermions.

The super Lie algebra associated with the $\pi$SUSY oscillator may be
considered in a more general context. This line of reasoning leads
to the introduction of supersymmetric paraquantum mechanics. The
defining superalgebra of SPQM determines the degeneracy structure of
the energy spectrum. The matrix reperesentation of the conserved charges
reveals the differences and the similarities between SPQM and
SQM. The Witten index can also be defined for SPQM. It possesses
the topological invariance property and signifies the breaking of
$\pi$SUSY, similarly to the ordinary SQM case.

The Hilbert space structure of the $\pi$SUSY oscillator is also
analyzed in detail. A remarkable observation is that the presence
of ($p=2$) parafermi operators does not lead to a triple grading of
the spectrum degeneracy. In fact, the general results obtained
in the context of supersymmetric paraquantum mechanics are shown to
be valid for the oscillator case. This serves as an independent check
on the results obtained in Sec.~6.

\section*{Acknowledgments}
I would like to thank Kamran Saririan, Bahman Darian, Ertu\"grul Demircan
and Stathis Tompaidis for mailing me some of the references.
I wish to also acknowledge Shahin Rouhani for his constructive comments.

\end{document}